\documentclass[aps,prl,showpacs,twocolumn,groupedaddress]{revtex4}

\usepackage{graphicx}
\usepackage{dcolumn}
\usepackage{bm}
\usepackage{color}

\begin{document}

\preprint{APS/123-QED}

\title{Breathing Pyrochlore Lattice Realized in $A$-Site Ordered Spinel Oxides LiGaCr$_4$O$_8$ and LiInCr$_4$O$_8$
}

\author{Yoshihiko Okamoto$^{1,*}$, 
G\o ran J. Nilsen$^{1,\dagger}$, J. Paul Attfield$^{2}$, and Zenji Hiroi$^{1}$}
\affiliation{
$^{1}$Institute for Solid State Physics, University of Tokyo, Kashiwa 277-8581, Japan\\
$^{2}$Centre for Science at Extreme Conditions and School of Chemistry, University of Edinburgh, Edinburgh EH9 3JZ, United Kingdom\\
}

\date{\today}

\begin{abstract}

A unique type of frustrated lattice is found in two $A$-site ordered spinel oxides, LiGaCr$_4$O$_8$ and LiInCr$_4$O$_8$. Because of the large size mismatch between Li$^+$ and Ga$^{3+}$/In$^{3+}$ ions at the $A$ site, the pyrochlore lattice, made up of Cr$^{3+}$ ions carrying spin 3/2, becomes an alternating array of small and large tetrahedra, i.e., a ``breathing'' pyrochlore lattice. We introduce a parameter, the breathing factor $B_{\mathrm{f}}$, which quantifies the degree of frustration in the pyrochlore lattice: $B_{\mathrm{f}}$ is defined as $J^{\prime}$/$J$, where $J^{\prime}$ and $J$ are nearest-neighbor magnetic interactions in the large and small tetrahedra, respectively. LiGaCr$_4$O$_8$ with $B_{\mathrm{f}}$ $\sim$ 0.6 shows magnetic susceptibility similar to that of conventional Cr spinel oxides such as ZnCr$_2$O$_4$. In contrast, LiInCr$_4$O$_8$ with a small $B_{\mathrm{f}}$ $\sim$ 0.1 exhibits a spin-gap behavior in its magnetic susceptibility, suggesting a proximity to an exotic singlet ground state. Magnetic long-range order occurs at 13.8 and 15.9 K for LiGaCr$_4$O$_8$ and LiInCr$_4$O$_8$, respectively, in both cases likely owing to the coupling to structural distortions. 
\end{abstract}

\pacs{Valid PACS appear here}
\maketitle

Transition metal oxides $AB_2$O$_4$ crystallizing in the spinel structure provide us with a rich playground for studying the physics of geometrical frustration. Transition metal $B$ atoms, which are octahedrally coordinated by oxide ions, form a three-dimensional network of tetrahedra, i.e., the pyrochlore lattice.
Various interesting phenomena have been observed arising from geometrical frustration concerning the spin and charge degrees of freedom on this lattice. Typical examples are the Verwey transition in Fe$_3$O$_4$~\cite{Verwey-1,Verwey-2}, a heavy-Fermion state in LiV$_2$O$_4$~\cite{LiV2O4}, and a heptamer formation in AlV$_2$O$_4$~\cite{AlV2O4}. 

$A$Cr$_2$O$_4$ with a nonmagnetic $A^{2+}$ ion, such as Zn$^{2+}$, Mg$^{2+}$, Cd$^{2+}$, or Hg$^{2+}$ at the tetrahedral site, and Cr$^{3+}$ ions at the octahedral site is of particular interest as a frustrated spin system~\cite{A-1}. It is a Mott insulator with three 3$d$ electrons localized at Cr$^{3+}$, yielding localized $S$ = 3/2 Heisenberg spin. Various magnitudes of antiferromagnetic interactions occur between nearest-neighbor spins, as evidenced by a range of negative Weiss temperatures of $-$390, $-$370, $-$70, and $-$32 K for $A$ = Zn, Mg, Cd, and Hg, respectively~\cite{Cd-1,Hg-1}. $A$Cr$_2$O$_4$ undergoes antiferromagnetic long-range order at 12, 12.4, 7.8, and 5.8 K, respectively~\cite{Cd-1,Hg-1,Mg-1}, which is accompanied by a lattice distortion which lowers the crystal symmetry~\cite{Mg-1,Zn-1,Cd-2}. Plausibly, there is an inherent structural instability in the spinel structure that can couple with the spin degree of freedom so as to lift the magnetic frustration.

\begin{figure}
\includegraphics[width=7.5cm]{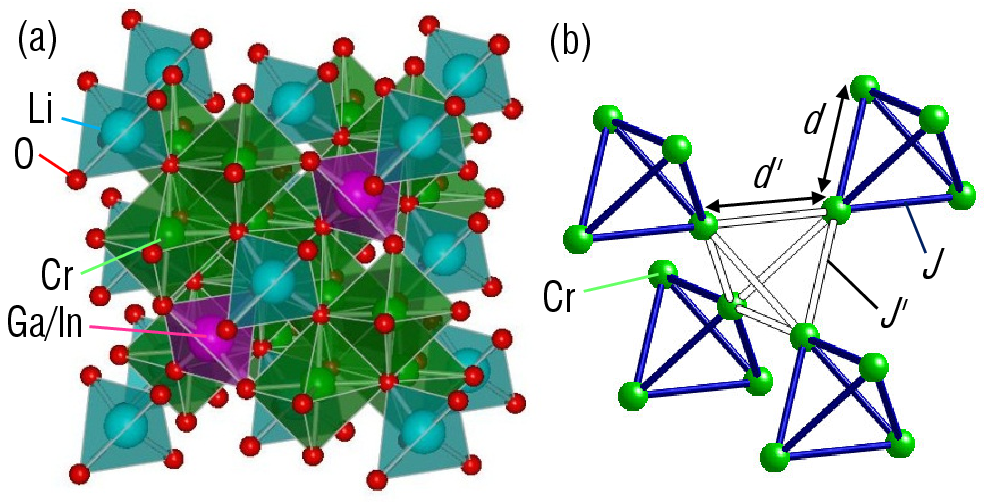}
\caption{\label{F1}
(color online) (a) Crystal structure of LiGaCr$_4$O$_8$ and LiInCr$_4$O$_8$. Coordination polyhedra made of oxide ions are depicted. 
(b) Breathing pyrochlore lattice made of Cr$^{3+}$ ions embedded in the two compounds. Cr-Cr bonds on the small (filled sticks) and large tetrahedra (open sticks) have bond lengths $d$ and $d^{\prime}$ and antiferromagnetic interactions $J$ and $J^{\prime}$, respectively.
}
\end{figure}

In this Letter, we study two spinel oxides, LiGaCr$_4$O$_8$ and LiInCr$_4$O$_8$, which both contain two metal ions at the $A$ site. Joubert and Durif prepared them in 1966~\cite{LiACr4O8} and found that they crystallize in a modified spinel structure with space group $F\bar{4}3m$, a subgroup of $Fd\bar{3}m$ for the conventional spinel oxides;
an inversion center at the octahedral site present in $Fd\bar{3}m$ is missing in $F\bar{4}3m$. A structural model was proposed in which Li and Ga/In atoms alternately occupy the tetrahedral sites, resulting in the zinc-blende type arrangement, although structural refinements were not performed~\cite{LiACr4O8}. This type of $A$-site order is likely, because it minimizes electrostatic energy arising from the large difference in the valence states between Li$^+$ and Ga$^{3+}$/In$^{3+}$. 

\begin{figure}
\includegraphics[width=8cm]{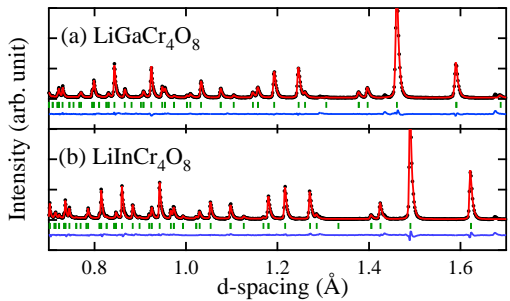}
\caption{\label{fig2} 
(color online) Powder neutron diffraction patterns for LiGaCr$_4$O$_8$ (a) and LiInCr$_4$O$_8$ (b) measured on bank 6 (2$\theta$ = 154$^{\circ}$) of the GEM diffractometer at room temperature. Filled circles are experimental data, and vertical bars indicate the positions of Bragg reflections. The curve on the data shows a calculated pattern, and the bottom curve shows a difference plot between the experimental and calculated intensities. 
}
\end{figure}

We are interested in the Cr pyrochlore lattices of these compounds, because the local chemical pressure caused by the difference in ionic radii of Li$^+$ and Ga$^{3+}$/In$^{3+}$ should result in the Cr$_4$ tetrahedra expanding and contracting alternately, whilst keeping their shapes regular, as shown in Fig. 1(b). We call this type of lattice the ``breathing'' pyrochlore lattice. The resulting modulation in bond lengths produces two kinds of nearest-neighbor magnetic interactions, $J$ and $J^{\prime}$, without relieving frustration. The spin Hamiltonian of the breathing pyrochlore lattice can thus be written as $H$ = $J$$\Sigma_{ij}\mathbf{S}_{i} \cdot \mathbf{S}_j$ $+$ $J^{\prime}$$\Sigma_{ij}\mathbf{S}_{i} \cdot \mathbf{S}_j$, where the summations over $ij$ in the first and second terms run over the Cr-Cr bonds of small and large tetrahedra, respectively; a uniform pyrochlore antiferromagnet is yielded when $J$ = $J^{\prime}$, and isolated tetramers are realized for $J^{\prime}$ = 0.

A theoretically predicted ground state for the uniform pyrochlore antiferromagnet is a spin liquid with a finite spin gap~\cite{Theory1,Theory2,Theory3}. This state has not yet been evidenced experimentally because actual compounds always suffer from various perturbations such as lattice deformation or defects. In the weak coupling limit, $J^{\prime}$ = 0, the ground state is also a gapped state, but with a considerably larger gap due to singlet formation on isolated tetrahedra. It would therefore be intriguing to examine how the two states are connected as a function of $B_{\mathrm{f}}$ = $J^{\prime}$/$J$ in the breathing pyrochlore lattice. LiGaCr$_4$O$_8$ and LiInCr$_4$O$_8$ are apparently the right compounds to study this issue.
The study on the breathing pyrochlore lattice would enable us to get a novel insight on the ground state of the uniform pyrochlore lattice as well as to explore a new phenomenon caused by the breathing. 

\begin{table}
\caption{\label{table1}Crystallographic parameters for LiGaCr$_4$O$_8$ and LiInCr$_4$O$_8$ (both $F\bar{4}3m$) determined by means of powder neutron diffraction. The lattice constant is $a$ = 8.2551(7) and 8.4205(5) \r{A}, respectively.
The range of $R$ factors obtained across all banks of data are given, and are found to be consistent with a Le Bail fit of only the main phase. $B$ is the thermal displacement parameter. The following constraints are assumed: $B$(Ga1) = $B$(Ga2), $B$(In1) = $B$(In2), and $B$(Li1) = $B$(Li2).}
\begin{ruledtabular}
\small
\begin{tabular}{lllllll}
&  & $x$ & $y$ & $z$ & Occ. & $B$ (\r{A}$^2$) \\
\hline
\multicolumn{7}{l}{LiGaCr$_4$O$_8$ ($R_{\mathrm{p}}$ = 3.72-4.77, $R_{\mathrm{wp}}$ = 4.79-6.49)} \\
Li1 & 4a & 0 & 0 & 0 & 0.994(7) & 1.96(24) \\
Ga1 & 4a & 0 & 0 & 0 & 0.006(7) & 0.53(4) \\
Li2 & 4d & 3/4 & 3/4 & 3/4 & 0.006(7) & 1.96(24) \\
Ga2 & 4d & 3/4 & 3/4 & 3/4 & 0.994(7) & 0.53(4) \\
Cr & 16e & 0.3728(3) & $x$ & $x$ & 1 & 0.33(3) \\
O1 & 16e & 0.13649(14) & $x$ & $x$ & 1 & 0.44(3) \\
O2 & 16e & 0.61889(13) & $x$ & $x$ & 1 & 0.37(3) \\
\hline
\multicolumn{7}{l}{LiInCr$_4$O$_8$ ($R_{\mathrm{p}}$ = 4.12-5.84, $R_{\mathrm{wp}}$ = 5.92-7.11)} \\
Li1 & 4a & 0 & 0 & 0 & 0.992(11) & 1.09(28) \\
In1 & 4a & 0 & 0 & 0 & 0.008(11) & 0.35(11) \\
Li2 & 4d & 3/4 & 3/4 & 3/4 & 0.008(11) & 1.09(28) \\
In2 & 4d & 3/4 & 3/4 & 3/4 & 0.992(11) & 0.35(11) \\
Cr & 16e & 0.3719(3) & $x$ & $x$ & 1 & 0.14(3) \\
O1 & 16e & 0.1377(2) & $x$ & $x$ & 1 & 0.38(4) \\
O2 & 16e & 0.61069(14) & $x$ & $x$ & 1 & 0.18(4) \\
\end{tabular}
\end{ruledtabular}
\end{table}

Polycrystalline samples of LiGaCr$_4$O$_8$ and LiInCr$_4$O$_8$ were prepared by the solid state reaction method. 
Li$_2$CO$_3$, Cr$_2$O$_3$, and Ga$_2$O$_3$/In$_2$O$_3$ powders were mixed in 1:4:1 molar ratio, then the mixture was sintered at 1000 $^{\circ}$C for one day and at 1100 $^{\circ}$C for another day.
Energy dispersive powder neutron diffraction experiments were carried out at room temperature on the General Materials (GEM) diffractometer at the ISIS pulsed neutron source. Rietveld analysis for the structural refinement was performed using the Fullprof program on 4 out of the 6 constant angle banks of experimental data using common structural parameters. Magnetic susceptibility and heat capacity measurements were performed in an MPMS and PPMS (both Quantum Design), respectively. 

Powder neutron diffraction patterns taken at room temperature for polycrystalline samples of LiGaCr$_4$O$_8$ and LiInCr$_4$O$_8$ are shown in Fig. 2. In addition to reflections expected for $Fd\bar{3}m$, forbidden reflections that are allowed for $F\bar{4}3m$, such as 002, are observed. 
Rietveld refinements using data from 4 banks were attempted starting from either case of full occupation of Li at the 4a site and Ga/In at the 4d site, as in the case of LiFe$_2$Rh$_3$O$_8$~\cite{LiFeRh}, or vice versa. Irrespective of the initial conditions, the refinements converged in a model in which the 4a and 4d sites are fully occupied by Li and Ga/In, respectively, within error bars. This is consistent with the model by Joubert and Durif~\cite{LiACr4O8}. Details of the refinement are given in Table I. The lattice parameters obtained are 8.2551(7) and 8.4205(5) \r{A}, respectively, which are similar to 8.243(3) and 8.411(3) \r{A} reported previously~\cite{LiACr4O8}. Thus, we have confirmed that the $F\bar{4}3m$ model with perfect $A$-site order is appropriate for the two compounds and successfully obtained reliable atomic coordinates. 

\begin{figure}
\includegraphics[width=9cm]{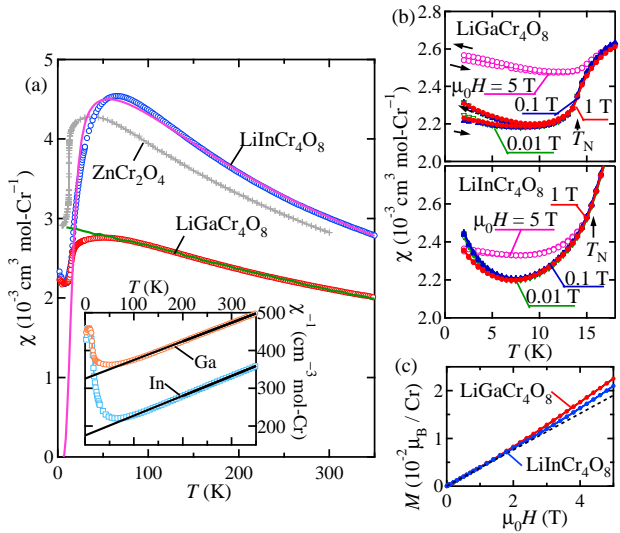}
\caption{\label{fig3} 
(color online) (a) Temperature dependence of magnetic susceptibility $\chi$ measured in a magnetic field of 1 T for polycrystalline samples of LiGaCr$_4$O$_8$ and LiInCr$_4$O$_8$. The data for a polycrystalline sample of ZnCr$_2$O$_4$ are also shown for comparison~\cite{Cd-1}. 
Solid lines are calculated curves produced by classical Monte Carlo simulations with $J^{\prime}$ = 0.5$J$ ($J$ = 53.3 K) for the Ga data and by the exact result for an isolated tetrahedron ($J$ = 56.8 K) for the In data.
The inset shows inverse susceptibilities with Curie-Weiss fits. (b) Temperature dependences of field-cooled and zero-field-cooled $\chi$'s measured at various magnetic fields
for LiGaCr$_4$O$_8$ (upper) and LiInCr$_4$O$_8$ (lower). 
(c) $M$-$H$ curves for LiGaCr$_4$O$_8$ and LiInCr$_4$O$_8$ measured at 5 K. The broken line shows an initial straight line.
}
\end{figure}

The zinc blende type order of small Li$^+$ and large Ga$^{3+}$/In$^{3+}$ ions on the 4a and 4d sites gives rise to a local chemical pressure on the Cr pyrochlore lattice, resulting in an alternation of the size of the Cr$_4$ tetrahedra. The Cr-Cr distances of the small and large tetrahedra, denoted as $d$ and $d^{\prime}$, are found to be 2.867(4) and 2.970(4) \r{A} for LiGaCr$_4$O$_8$, and 2.903(4) and 3.052(4) \r{A} for LiInCr$_4$O$_8$, respectively. The differences between $d$ and $d^{\prime}$ are 3.5\% and 4.9\%, respectively, meaning a stronger alternation or breathing in LiInCr$_4$O$_8$ than in LiGaCr$_4$O$_8$. 

Figure 3(a) shows the temperature dependence of magnetic susceptibility $\chi$ for LiGaCr$_4$O$_8$ and LiInCr$_4$O$_8$. The inverse of $\chi$, shown in the inset, exhibits a linear temperature dependence above $\sim$100 K, following the Curie-Weiss law $\chi$ = $C$/($T - \theta_{\mathrm{W}}$), where $C$ and $\theta_{\mathrm{W}}$ are the Curie constant and the Weiss temperature; a fit to the data between 200 and 350 K yields $C$ = 2.025(3) cm$^3$ K mol-Cr$^{-1}$ and $\theta_{\mathrm{W}}$ = $-$658.8(4) K for LiGaCr$_4$O$_8$, and $C$ = 1.899(4) cm$^3$ K mol-Cr$^{-1}$ and $\theta_{\mathrm{W}}$ = $-$331.9(4) K for LiInCr$_4$O$_8$. The values of $C$ correspond to effective moments of $\mu_{\mathrm{eff}}$ = 4.024 $\mu_{\mathrm{B}}$ and 3.897 $\mu_{\mathrm{B}}$ per Cr atom, or Lande $g$-factors of $g$ = 2.078 and 2.012 for $S$ = 3/2, respectively. The large and negative $\theta_{\mathrm{W}}$ indicates that average magnetic interactions 
are strongly antiferromagnetic. 

The $\chi$ versus $T$ curve of LiGaCr$_4$O$_8$ resembles that of ZnCr$_2$O$_4$, as indicated in Fig. 3(a). It 
shows a broad peak at $\sim$45 K, which represents a development of antiferromagnetic short-range order. In the case of ZnCr$_2$O$_4$, the corresponding broad peak appears at $\sim$30 K, reflecting stronger magnetic interactions in LiGaCr$_4$O$_8$. On further cooling, $\chi$ decreases steeply at $\sim$14 K, where long-range order sets in, as also evidenced by a sharp peak in heat capacity at $T_{\mathrm{N}}$ = 13.8 K (Fig. 4), close to that of ZnCr$_2$O$_4$. Therefore, the breathing of the pyrochlore lattice in LiGaCr$_4$O$_8$ has little influence on the magnetic properties. 

\begin{figure}
\includegraphics[width=6.2cm]{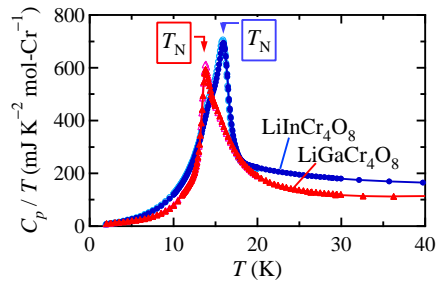}
\caption{\label{fig4} 
(color online) Temperature dependence of heat capacity divided by temperature $C_p$/$T$ for polycrystalline samples of LiGaCr$_4$O$_8$ and LiInCr$_4$O$_8$. Filled and open symbols represent data measured at magnetic fields of 0 and 9 T for each sample, respectively. 
}
\end{figure}

In contrast, the $\chi$ of LiInCr$_4$O$_8$ shows a temperature dependence which is obviously distinguishable from that of the Ga analogue: it rapidly decreases with decreasing temperature below 65 K, which is reminiscent of the opening of a spin gap. 
The gap size is roughly estimated to be $\Delta$ = 56.8(2) K by a fit to the exact result for an isolated tetrahedron of $S$ = 3/2, where the diamagnetic contribution of core electrons of $-$4.0 $\times$ 10$^{-5}$ cm$^3$ mol-Cr$^{-1}$ is included~\cite{dia}, as shown in Fig. 3(a). At yet lower temperature, a sharp peak in $C_p$/$T$ (Fig. 4) indicates long-range order at $T_{\mathrm{N}}$ = 15.9 K. Based on this, we believe the ground state of LiInCr$_4$O$_8$ to be in proximity to a spin-gapped state, despite the long-range order at low temperature. As shown in Fig. 3(b), the $\chi$ of each compound is nearly insensitive to magnetic fields below 1 T, beyond which it suddenly increases at 5 T below $T_{\mathrm{N}}$. This is clearly observed in the $M$-$H$ curves shown in Fig. 3(c), where an upturn from the initial straight line appears at approximately 2 T in each compound. This may be due to a spin flop transition in the ordered state. 

\begin{figure}
\includegraphics[width=6.5cm]{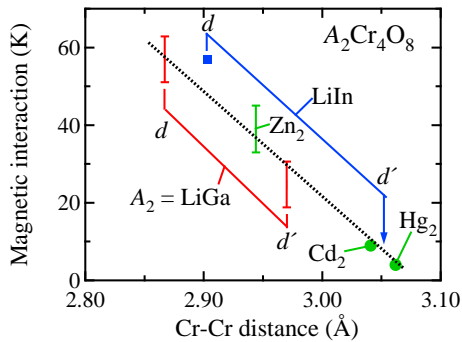}
\caption{\label{fig5} 
(color online) Nearest-neighbor magnetic interaction versus the Cr-Cr distance for various Cr spinel oxides. The $J$ value of ZnCr$_2$O$_4$ are estimated by fitting $\chi$, $C_p$, and EPR intensity~\cite{Zn-5,Zn-6,Zn-7}.
The $J$'s of CdCr$_2$O$_4$ and HgCr$_2$O$_4$ are estimated by Curie-Weiss fits to $\chi$~\cite{Cd-1,Hg-1,Hg-3}. The dotted line is a guide for the eyes. Plotted for LiGaCr$_4$O$_8$ or LiInCr$_4$O$_8$ is a pair of points corresponding to the two Cr-Cr distances in the small and large tetrahedra. 
}
\end{figure}

We now consider magnetic interactions in the Cr spinel oxides. Generally, it is well established that the shorter the Cr-Cr distance, the stronger the antiferromagnetic interaction, as shown in Fig. 5, because exchange interactions are dominantly mediated by direct overlap between Cr $t_{\mathrm{2g}}$ orbitals along the Cr-Cr bond~\cite{A-2,A-3}. For example, the $J$ of ZnCr$_2$O$_4$ is 33-45 K~\cite{Zn-5,Zn-6,Zn-7}, almost ten times larger than 4 K for HgCr$_2$O$_4$~\cite{Hg-1,Hg-3}, which comes from a 4\% difference in the Cr-Cr distance between ZnCr$_2$O$_4$ (2.944 \r{A}) and HgCr$_2$O$_4$ (3.062 \r{A}). Interestingly, there is an approximate linear relationship between $J$ and the bond length for $A$Cr$_2$O$_4$, as indicated by the dotted line in Fig. 5.

A remarkable feature of the present Ga and In compounds is the extremely short Cr-Cr distance within the small Cr$_4$ tetrahedra compared with that of $A$Cr$_2$O$_4$: $d$ = 2.867 \r{A} (Ga) and 2.903 \r{A} (In) are much shorter than 2.944 \r{A} in ZnCr$_2$O$_4$~\cite{Zn-4}, suggesting that $J$ in the Ga and In compounds are larger than $J$ = 33-45 K in ZnCr$_2$O$_4$. On the other hand, the distance $d^{\prime}$ = 2.970 \r{A} within the large tetrahedra of LiGaCr$_4$O$_8$ is comparable to the Cr-Cr distance of ZnCr$_2$O$_4$, while the $d^{\prime}$ = 3.052 \r{A} for LiInCr$_4$O$_8$ lies between 3.041 \r{A} in CdCr$_2$O$_4$ and 3.062 \r{A} in HgCr$_2$O$_4$. Assuming that the linear relationship observed in $A$Cr$_2$O$_4$ holds in LiGaCr$_4$O$_8$ and LiInCr$_4$O$_8$, we arrive at estimates of $J \sim$ 60 and 50 K and $J^{\prime} \sim$ 30 and 6 K, which give a breathing factor $B_{\mathrm{f}}$ = $J^{\prime}$/$J \sim$ 0.5 and 0.12, respectively. 

To more accurately estimate $J$ and $J^{\prime}$, we carried out fitting of the $\chi$ data shown in Fig. 3(a). Classical Monte Carlo simulations were performed using the spinmc program of the ALPS package (1024 sites, periodic boundary conditions). Although it was difficult to determine the two values uniquely, we could obtain $J$ for various $B_{\mathrm{f}}$'s, e.g. $J$ = 62.9(5), 53.3(1), and 51.0(3) K for $B_{\mathrm{f}}$ = 0.3, 0.5, and 0.6, respectively; a typical fitting curve for $B_{\mathrm{f}}$ = 0.5 is shown in Fig. 3(a). Taking into account of $J \sim$ 50 K from the universal line in Fig. 5, we decide $B_{\mathrm{f}}$ = 0.6 for LiGaCr$_4$O$_8$. 

In contrast to LiGaCr$_4$O$_8$, the fit to the Monte Carlo results was poor for LiInCr$_4$O$_8$. 
Thus, we adopt $J$ = 56.8(2) K from the fitting to the exact result mentioned before,
which lies close to the universal line in Fig. 5. 
We also tried to fit the data by taking account of $J^{\prime}$ in the mean-field approximation. However, the improvement of fitting was limited, and it resulted in an unreasonably large ferromagnetic $J^{\prime}$. Thus, we assume $J^{\prime} \sim$ 6 K from the universal relation in Fig. 5, which gives $B_{\mathrm{f}} \sim$ 0.1 for LiInCr$_4$O$_8$.
The breathing factor can be an important parameter to tune the ground state of pyrochlore lattice antiferromagnets. 

Finally, we describe the characteristics of the ordering transitions in the two compounds. Their $C_p$/$T$ curves, shown in Fig. 4, exhibit sharp peaks at 13.8 and 15.9 K, respectively, clear evidence for magnetic transitions. Interestingly, the peak shape for LiGaCr$_4$O$_8$ is apparently different from that of a conventional second-order magnetic transition: the $C_p$/$T$ shows a broad shoulder at $\sim$16 K above $T_{\mathrm{N}}$ and gradually decreases with increasing temperature, indicating that a large spin entropy is retained above $T_{\mathrm{N}}$ due to the development of antiferromagnetic short-range order. 
In LiInCr$_4$O$_8$, there is also a large entropy release above $T_{\mathrm{N}}$, associated with a spin-gap formation. 
These magnetic transitions are quite robust against magnetic fields: the two $C_p$/$T$ curves measured at 0 and 9 T nearly overlap to each other in each compound.

The origin of long-range order in the present two compounds is not perfectly confirmed. In the absence of structural distortions, geometrical frustration should remain and favor a spin-liquid ground state. In $A$Cr$_2$O$_4$, a structural transition always takes place simultaneously with the magnetic transition. The similar $T_{\mathrm{N}}$'s of 5-15 K, in spite of the broad variation of $J$, suggest a common structural instability that is coupled with spins through strong spin-lattice interactions. This is also likely the case for the present compounds. We plan to investigate the crystal and magnetic structures across $T_{\mathrm{N}}$ by means of NMR, X-ray, and neutron diffraction experiments. Moreover, inelastic neutron scattering experiments will aid in establishing the presence (or absence) of a gap in either material, as well as allowing for quantitative determination of $J$ and $J^{\prime}$. Although the compounds assume long-range ordered ground states, fingerprints of neighboring spin liquid states or the effect of frustration may be observable as in Ref. 24 and 25.

In summary, two spinel oxides LiGaCr$_4$O$_8$ and LiInCr$_4$O$_8$ with the tetrahedral sites alternately occupied by Li$^+$ and Ga$^{3+}$/In$^{3+}$ ions are found to be unique frustrated antiferromagnets with breathing pyrochlore lattices. LiGaCr$_4$O$_8$, with a lesser degree of breathing, shows similar magnetic properties to the conventional Cr spinel oxides with a uniform pyrochlore lattice, while LiInCr$_4$O$_8$ shows a spin-gap behavior caused by a large alternation of magnetic interactions in the more breathing pyrochlore lattice. The breathing of the pyrochlore lattice appears to be an important parameter to explore interesting phenomena in frustrated magnets. 

We thank H. Tsunetsugu, M. Isobe, H. Ueda, and T. Nakazono for helpful discussion.


\end{document}